*Tunable Intracavity Coherent Up-conversion with Giant Nonlinearity*
*in a Polar Fluidic Medium*

Daichi Okada*, Hiroya Nishikawa and Fumito Araoka*

RIKEN Center for Emergent Matter Science (CEMS), 2-1 Hirosawa, Wako, Saitama 351-0198, Japan

Present address (Dr. Daichi Okada): Faculty of Electrical Engineering and Electronics, Kyoto Institute of Technology, Matsugasaki, Sakyo-ku, Kyoto, 606-8585, Japan

E-mail: oka-d@kit.ac.jp; fumito.araoka@riken.jp

Keywords: Second harmonic generation • Polar nematic liquid crystal • Optical microcavity • Up-conversion • Nonlinear optical response

**Abstract**

We demonstrate a novel microcavity-based photon up-conversion using second harmonic generation (SHG) from a polar nematic fluid media doped with a laser dye. The present idea is based on coherent light generation via simultaneous frequency doubling and stimulated emission (lasing) inside a microcavity. The polar nematic fluid equips very high even-order optical nonlinearity due to the polar symmetry and large dipole moment along the molecular long axis. At the same time, its inherent fluidic nature allows us to easily functionalize the media just by doping, in the present case, with an emissive laser dye. Our demonstrated system exhibits a giant nonlinear optical response to input light, while enabling spectral narrowing and multiple-signal output of up-converted light, that is not attainable though the simple SH-conversion of input light. Furthermore, susceptibility of the liquid crystal offers dynamic modulation capabilities under external stimulus, such as signal switching with electric field application or wavelength tuning through temperature variation. Such a brand-new type of simple coherent flexible up-conversion system must be promising as a new principle for easy-accessible and down-scalable wavelength conversion devices.

Photon up-conversion, which is a process to produce higher-energy photons from lower-energy photons, is not just useful for wavelength conversion in laser systems but also a candidate technology to boost up the efficiency of important photo-processes for photovoltaics, photocatalysis and optogenetics. Second harmonic generation (SHG) is one of the most well-known and fundamental photon up-conversion processes via the optical nonlinearity exhibited in symmetry-broken materials, which is ubiquitous and hence utilized in various applications such as bio-imaging, biotherapy, optical communication, optical computing, etc. However, generally the intrinsic conversion efficiency of SHG is so small, it is necessary to introduce extrinsic mechanism to increase the light-matter interactions, such as by phase matching [1], photonic effect [2], or confinement in resonators like Fabry-Pérot etalons, micro/nano-spheres/disks, plasmonic nano structures [3]. Among those, one of the most common approaches would be the use of an optical microcavity structure, as a convenient and effective method to obtain sufficiently high conversion efficiency in SHG. Since the photons are densely confined in a small volume with a size comparable with the optical wavelength, light-matter interaction can be strongly enhanced and provide one to three-orders of magnitude stronger SHG [3]. This paper demonstrates a novel type of photon up-conversion using a Fabry-Pérot microcavity, in which dye-doped polar nematic liquid crystal (PNLC) enables two-step intra-cavity coherent light generation through stimulated emission (lasing) and simultaneous frequency doubling (Figure 1). The main role of the PNLC is wavelength conversion due to its excellent nonlinear optical property originated from the polar symmetry and large dipole moment along the molecular long axis. At the same time, its inherent fluidic nature allows itself to equip with an optical gain just by doping with a laser dye. Another advantage of the system is easy-preparation. Since the F-P cavity structure is essentially the same as a sandwich-type

LC cell, the LC material can be easily introduced in the FP cavity again with the aid of the fluidic nature. As an optical device, it can easily improve the monochromaticity of the fundamental light and the resulted SHG efficiency, in addition to the giant nonlinear optical response to the incident light. Multi-mode signal output is also achievable depending on the cavity condition. Such an advantageous characteristics are not easily attainable in the conventional SHG system. Furthermore, the flexibility of LC provides our optical device with switchability and tunability in response to external fields. Our demonstrated coherent up-conversion system mediated both stimulated emission and SH conversion holds promise for future applications in information processing devices based on nonlinear optics.

As a PNLC medium, the recently-found ferroelectric nematic liquid crystal is used to prove our concept [4]. Although this kind of liquid crystal materials have a highly fluidic nature, they still exhibit extremely strong SHG (approximately 70 times higher compared to a quartz reference), elucidating a remarkable nonlinear optical property [4d]. Meanwhile, doping a small molecule is an effective way to functionalize a liquid crystal material. For example, mixing a liquid crystal and a photo-isomerizable azobenzene causes photo-induced athermal phase transition [4g, 5], and simple doping of a cholesteric liquid crystal with an organic laser dye makes possible cavity-less lasing [6]. Indeed, this kind of approach also works for PNLC, and already several examples have been demonstrated to fuse additional functional properties into ferroelectricity to realize unprecedented applications, such as photo-variable capacitors [4h], tunable polar cholesteric reflector [4c, g], electrically controllable microlenses [4o], enhanced nonlinear optical response [4k]. Thus, we hit an idea of a flexible coherent up-conversion system using a doped PNLC as a functionalized nonlinear optical medium. The PNLC, RM734, was doped with a boron difluoride curcuminoid-based fluorescent dye (1 wt%) (Figure 2a), which has the capability of lasing in the near-infrared range (700~800 nm) [7]. Since the primary absorption band of PNLC is in the UV region below 350 nm, the self-absorption effect is negligible in any of the following result (Figure S1). The doped PNLC was injected into a microcavity structure consisting of two quartz substrates separated with a 5 μm spacer (Figure S2), inner surfaces of which were furnished with dielectric reflectors (%T< 0.5 %, covering 700~800 nm) (Figure S3). To obtain uniform planar orientation, rubbed polyimide was used for alignment layers. The nice unidirectional alignment was confirmed by polarized optical microscopy (Figure S4). The home-built optical setup is schematized in Figure S5. Upon irradiation of the fundamental/excitation laser pulses (~200 fs, 10 kHz) from an optical parametric amplifier, the output light was spectroscopically recorded by an optical multi-channel analyser (OMA), and simultaneously, its intensity was measured by a photon-counter attached to a monochromator.

Firstly, we examined the conventional transmission SHG from the doped PNLC introduced in a usual liquid crystal cell without reflectors. Heated up to the isotropic temperature at 180 ºC, the sample was gently cooled down upon monitoring the output light. After undergoing the phase transition to the ferroelectric nematic liquid crystal phase at about 120 ºC, a strong monochromatic light peak appears at the half wavelength (~400 nm) of the fundamental light (~800 nm) (Figure 2b). Hereafter, the temperature was fixed at 120 ºC, otherwise noted. The obtained signal intensity nearly-quadratically increases with the fundamental light intensity (Figure 2c), clearly evidencing SHG even in the doped PNLC (The reason of the slightly-lower logarithmic slope (1.84) than the ideal value (2) could be the self-absorption effect of the dye). Due to the self-absorption effect of doped dye, the SHG intensity is slightly reduced in comparison with the RM734 itself (Figure 2b, inset). We scanned the fundamental light wavelength in the lasing range of the dye (from 700 to 800 nm) and found that there is an average loss of SHG of about 30% in comparison to the undoped PNLC (Figure S6). This could be due to the self-absorption effect and/or the slight disordering in the doped PNLC but still not so significant in our present demonstration. Note that the output SHG is highly linearly-polarized to the rubbed direction (Figure 2d), so that the disorder effect should be minor.

Next, we investigated the lasing and up-conversion properties in the dye doped PNLC confined between the microcavity structure. To excite the laser dye, the irradiated laser wavelength was tuned to 600 nm, close to its absorption peak top. The beam spot was focused to ~100 um by a lens. Shown in Figure 3a and 3e are two different PL spectra for two different cavity conditions (Sample 1 and 2), originated from the deviation of the refractive index caused by slight disorder in the alignment of the liquid crystalline media or the inhomogeneity of the cavity length (cell thickness). As increasing the excitation intensity, narrow lasing peaks appeared at 777 nm (Sample 1) and 747 nm (Sample 2), and their intensity increased with obvious nonlinearity (Figure 3c, g, black dotted line). The difference of the lasing wavelengths in two samples is due to the above difference of the cavity conditions. The lasing threshold was estimated from the intersection of PL and the background (BG) slopes in the logarithmic plot (Figure 3c and 3g), as 0.23 nJ/pulse for Sample 1 and 1.75 nJ/pulse for Sample 2, and their nonlinearity below the saturation was evaluated as logarithmic slopes, 3.68 and 5.06, respectively (black dotted lines in Figure 3c and 3g). The difference of the threshold stems mainly from the difference in optical gain. Specifically, 777 nm is distant from the absorption region of the dye, so that less self-absorption in this region leads to a better optical gain than 747 nm.

In addition to lasing, a sharp peak appeared in the near ultra-violet region, at almost the half wavelength of the above lasing peak (Figure 3b and 3f). This additional emission peak also has almost the same excitation threshold as that of the lasing, and no emission can be detected below this threshold. Since this wavelength region is outside the reflection band of the dielectric reflectors of the microcavity, this peak is considered to be of the simultaneous SH conversion of the lasing light, that is, the coherent up-converted light. This up-converted light intensity also nonlinearly depends on the excitation intensity (red dotted line in Figure 3c and 3g), but its power law looks different. As estimated from Figure 3c and 3g, the logarithmic slopes for these two conditions are 9.75 and 12.30, respectively, much larger than those of the lasing itself. Such a giant nonlinearity is attributed to the combined nonlinearities of the simultaneous lasing and SHG processes in the present up-conversion system. By plotting the up-converted intensity versus the lasing intensity, one can see almost-quadratic dependencies with their logarithmic slopes of 2.55 and 2.31, as shown in Figure 3d and 3h, respectively. Of course, if this is the usual SHG, the logarithmic slope should be just 2 or slightly less, like in the case of Figure 2d. However, the slightly larger logarithmic slope means the higher efficiency than the usual conversion. Since the lasing acting as the fundamental light is a result of the confinement in the cavity while SHG not, we need to know at least the actual intensity of the lasing light field in the cavity to understand fully the present up-conversion behaviour. If there is any other effect such as photo-refractive effect or photo-bleaching effect, the situation is further complicated. The former may cause the power-dependent phase matching/unmatching between the lasing and SHG lights, which may change the efficiency. Anyway, the situation is too complicated to be analysed, and the full explanation will be kept for future problem.

      The system also enables us to modulate the up-conversion efficiency just by changing the optical condition in the cavity. It is known that in many cases, anisotropic dye molecules uniaxially align their molecular long axis to the nematic director, when embedded in a liquid crystal medium. Then, the transition dipole moment of the dye is also aligned, and this makes lasing most efficient with the use of an excitation light linearly polarized to the aligned direction of the liquid crystal. In the present case, the nonlinear polarization also directs along the director of PNLC. Thus, the most efficient up-conversion is achieved when excited in the other words, the up-conversion efficiency is modulated by the polarization state. Similarly, the output up-converted light is also linearly polarized along the alignment direction (Figure S7). The estimated dichroic ratio is about 0.78, so that the conversion efficiency is lower if the sample is excited with an excitation light orthogonal to the aligned direction.

Figure 4a compares the power laws in the present up-conversion system (600 nm→390 nm) and the usual SHG (800 nm→400 nm). Due to the high slope efficiency, it is evident that the former takes over the latter in the high intensity region. However, due to the saturation of optical pumping, the up-conversion intensity also saturates at high pump fluences, while SHG doesn't. Another notable feature is narrow bandwidth of the up-converted light. Generally, the character of the output SHG is succeeded from that of the fundamental light. For example, a broadband fundamental light of ultra-short pulses (typically, from a femto-second laser) results in broadband SHG. However, although the present up-conversion is based on SHG, narrower bandwidth (more monochromatic) output is always obtained independently from the bandwidth of the irradiated excitation light. This is because, the present up-converted light is generated though the intra-cavity SHG of the lasing, whose bandwidth is determined typically by the characteristics of the microcavity and the laser dye (Figure 4b).

Regarding the above, we introduce an interesting feature, to say, multi-mode coherent up-conversion. Because the lasing condition is influenced by the optical characteristic of the microcavity as mentioned above, we can realize multimode lasing by introducing spatial inhomogeneity in the system. Here, we demonstrate two examples of multi-mode lasing (red curves, Figure 4c), both of which are observed in a same disordered PNLC device but at different positions. Accordingly, the resulted up-converted light from each position also becomes multi-mode (blue curves, Figure 4c), obtaining a coherent up-converted light with multiple wavelength peaks despite the use of a single F-P device and a single excitation light of 600 nm.

Finally, we would like to demonstrate stimuli-responsive modulation under electric-field application and temperature variation. In order to apply an electric field, we fabricated a F-P microcavity device with dielectric reflectors, on the top of which transparent electrodes of indium-tin-oxide (ITO) are coated, as depicted in Figure 5a. Initially, a uniform texture was observed under polarization microscopy. Upon applying an AC electric field perpendicular to the substrate (20Vpp, 200Hz), dynamic and chaotic turbulent occurs and disrupts the ordered structure of PNLC (Figure 5a). As a consequence, lasing threshold hugely increases and leads to deactivation (Figures 5b, 5c, S8). Thus, the up-conversion is switched off by turning on the field application. When the electric field is turned off, the alignment restored and the up-conversion reappear. In such a way, we can switch on and off the up-conversion at will by the electric field application (Figure 5a, 5b, 5c). In addition, it is also possible to tune the wavelength of the up-converted light by heating or cooling through the continuous change in the cavity condition due to the temperature dependent refractive index and/or thermal expansion/shrinkage. As decreasing the temperature from 120ºC to 105ºC within the ferroelectric nematic phase, the lasing wavelength continuously blue-shifted, and accordingly, the up-conversion wavelength also blue-shifted (Figure 5d, 5e). These unprecedented properties further represent the uniqueness of our flexible coherent up-conversion system.

In this paper, we proposed the novel intracavity up-conversion system using a Fabry-Pérot microcavity filled with a dye-doped PNLC. The system is quite simple, yet provide us multitude of advantages, such as giant nonlinearity, increased efficiency, spectrum narrowing and generation of coherent multi-mode outputs. These characteristics are usually difficult to achieve in conventional SHG. In addition, since our system is inherently soft due to liquid crystallinity, the obtained signal further dynamically modulated by external stimuli. Such a flexible microcavity based coherent up-conversion system extends a usefulness of soft ferroelectrics and lead to the development of innovative future nonlinear optical devices.

**Acknowledgements**

This work was partly supported by Grant-in-Aid for Scientific research (B) (JP21H01801, JP23H01942), Grant-in-Aid for Challenging Research (Pioneering) (23K17341), and Young Scientists (JP22K14594, JP21K14605, 19K15438) from Japan Society for the Promotion of

**References**

[1] a) T. Stolt, J. Kim, S. Héron, A. Vesala, Y. Yang, J. Mun, M. Kim, Mikko. J. Huttunen, R. Czaplicki, M. Kauranen, J. Rho, P. Genevet, Phys. Rev. Lett. 2021, 126, 033901; b) X. Xu, C. Trovatello, F. Mooshammer, Y. Shao, S. Zhang, K. Yao, D. N. Basov, G. Cerullo, P. J. Schuck, Nat. Photon. 2022, 16, 698–706; c) Z. Li, W. Jin, F. Zhang, Z. Yang, S. Pan, ACS Cent. Sci. 2022, 8, 1557–1564; d) M. Mutailipu, J. Han, Z. Li, F. Li, J. Li, F. Zhang, X. Long, Z. Yang, S. Pan, Nat. Photon. 2023, 17, 694–701.; e) H. Hong, C. Huang, C. Ma, J. Qi, X. Shi, C. Liu, S. Wu, Z. Sun, E. Wang, K. Liu, Phys. Rev. Lett. 2023, 131, 233801; f) Y. Song, H. Yu, B. Li, X. Li, Y. Zhou, Y. Li, C. He, G. Zhang, J. Luo, S. Zhao, Adv Funct Materials 2024, 34, 2310407.

[2] a) A. A. Fedyanin, O. A. Aktsipetrov, D. A. Kurdyukov, V. G. Golubev, M. Inoue, Applied Physics Letters 2005, 87, 151111; b) K. I. Zaytsev, S. O. Yurchenko, Applied Physics Letters 2014, 105, 051902; c) D. Wei, C. Wang, H. Wang, X. Hu, D. Wei, X. Fang, Y. Zhang, D. Wu, Y. Hu, J. Li, S. Zhu, M. Xiao, Nature Photon 2018, 12, 596–600; d) N. Bernhardt, K. Koshelev, S. J. U. White, K. W. C. Meng, J. E. Fröch, S. Kim, T. T. Tran, D.-Y. Choi, Y. Kivshar, A. S. Solntsev, Nano Lett. 2020, 20, 5309–5314; e) Z. Liu, J. Wang, B. Chen, Y. Wei, W. Liu, J. Liu, Nano Lett. 2021, 21, 7405–7410; f) X. Zhao, J. Zhou, J. Li, J. Kougo, Z. Wan, M. Huang, S. Aya, Proc. Natl. Acad. Sci. U.S.A. 2021, 118, e2111101118; g) X. Zhao, H. Long, H. Xu, J. Kougo, R. Xia, J. Li, M. Huang, S. Aya, Proc. Natl. Acad. Sci. U.S.A. 2022, 119, e2205636119.

[3] a) Y. Pu, R. Grange, C.-L. Hsieh, D. Psaltis, Phys. Rev. Lett. 2010, 104, 207402; b) P. S. Kuo, J. Bravo-Abad, G. S. Solomon, Nat Commun 2014, 5, 3109; c) M.-L. Ren, W. Liu, C. O. Aspetti, L. Sun, R. Agarwal, Nat Commun 2014, 5, 5432; d) T. Chervy, J. Xu, Y. Duan, C. Wang, L. Mager, M. Frerejean, J. A. W. Münninghoff, P. Tinnemans, J. A. Hutchison, C. Genet, A. E. Rowan, T. Rasing, T. W. Ebbesen, Nano Lett. 2016, 16, 7352–7356; e) F. Yi, M. Ren, J. C. Reed, H. Zhu, J. Hou, C. H. Naylor, A. T. C. Johnson, R. Agarwal, E. Cubukcu, Nano Lett. 2016, 16, 1631–1636; f) T. Fryett, A. Zhan, A. Majumdar, Nanophotonics 2017, 7, 355–370; g) Q. Ai, F. Sterl, H. Zhang, J. Wang, H. Giessen, ACS Nano 2021, 15, 19409–19417; h) J. Shi, X. Wu, K. Wu, S. Zhang, X. Sui, W. Du, S. Yue, Y. Liang, C. Jiang, Z. Wang, W. Wang, L. Liu, B. Wu, Q. Zhang, Y. Huang, C.-W. Qiu, X. Liu, ACS Nano 2022, 16, 13933–13941; i) J.

Shi, X. He, W. Chen, Y. Li, M. Kang, Y. Cai, H. Xu, Nano Lett. 2022, 22, 688–694; j) R. Biswas, A. Prosad, L. A. S. Krishna, S. Menon, V. Raghunathan, Nanophotonics 2023, 12, 29–42; k) L. Qu, Z. Gu, C. Li, Y. Qin, Y. Zhang, D. Zhang, J. Zhao, Q. Liu, C. Jin, L. Wang, W. Wu, W. Cai, H. Liu, M. Ren, J. Xu, Adv Funct Materials 2023, 33, 2308484.

[4] a) H. Nishikawa, K. Shiroshita, H. Higuchi, Y. Okumura, Y. Haseba, S. Yamamoto, K. Sago, H. Kikuchi, Advanced Materials 2017, 29, 1702354; b) X. Chen, E. Korblova, D. Dong, X. Wei, R. Shao, L. Radzihovsky, M. A. Glaser, J. E. Maclennan, D. Bedrov, D. M. Walba, N. A. Clark, Proc. Natl. Acad. Sci. U.S.A. 2020, 117, 14021–14031; c) H. Nishikawa, F. Araoka, Advanced Materials 2021, 33, 2101305; d) J. Li, H. Nishikawa, J. Kougo, J. Zhou, S. Dai, W. Tang, X. Zhao, Y. Hisai, M. Huang, S. Aya, Sci. Adv. 2021, 7, eabf5047; e) R. J. Mandle, N. Sebastián, J. Martinez-Perdiguero, A. Mertelj, Nat Commun 2021, 12, 4962; f) X. Chen, E. Korblova, M. A. Glaser, J. E. Maclennan, D. M. Walba, N. A. Clark, Proc. Natl. Acad. Sci. U.S.A. 2021, 118, e2104092118; g) C. Feng, R. Saha, E. Korblova, D. Walba, S. N. Sprunt, A. Jákli, Advanced Optical Materials 2021, 9, 2101230; h) H. Nishikawa, K. Sano, F. Araoka, Nat Commun 2022, 13, 1142; i) R. J. Mandle, Soft Matter 2022, 18, 5014–5020; j) H. Nishikawa, K. Sano, S. Kurihara, G. Watanabe, A. Nihonyanagi, B. Dhara, F. Araoka, Commun Mater 2022, 3, 89; k) R. Xia, X. Zhao, J. Li, H. Lei, Y. Song, W. Peng, X. Zhang, S. Aya, M. Huang, J. Mater. Chem. C 2023, 11, 10905–10910; l) M. T. Máthé, B. Farkas, L. Péter, Á. Buka, A. Jákli, P. Salamon, Sci Rep 2023, 13, 6981; m) J. Szydlowska, P. Majewski, M. Čepič, N. Vaupotič, P. Rybak, C. T. Imrie, R. Walker, E. Cruickshank, J. M. D. Storey, P. Damian, E. Gorecka, Phys. Rev. Lett. 2023, 130, 216802; n) S. Marni, G. Nava, R. Barboza, T. G. Bellini, L. Lucchetti, Advanced Materials 2023, 35, 2212067; o) K. Perera, N. Haputhantrige, M. S. H. Himel, M. Mostafa, A. Adaka, E. K. Mann, O. D. Lavrentovich, A. Jákli, Advanced Optical Materials 2024, 12, 2302500; p) M. T. Máthé, K. Perera, Á. Buka, P. Salamon, A. Jákli, Advanced Science 2024, 11, 2305950.

[5] a) J.-H. Sung, S. Hirano, O. Tsutsumi, A. Kanazawa, T. Shiono, T. Ikeda, Chem. Mater. 2002, 14, 385–391; b) K. Matczyszyn, J. Sworakowski, J. Phys. Chem. B 2003, 107, 6039–6045; c) H. Yu, T. Ikeda, Advanced Materials 2011, 23, 2149–2180; d) H. K. Bisoyi, Q. Li, Chem. Rev. 2016, 116, 15089–15166; e) M. J. Moran, M. Magrini, D. M. Walba, I. Aprahamian, J. Am. Chem. Soc. 2018, 140, 13623–13627.

[6] a) A. Chanishvili, G. Chilaya, G. Petriashvili, R. Barberi, R. Bartolino, G. Cipparrone, A. Mazzulla, L. Oriol, Applied Physics Letters 2003, 83, 5353–5355; b) T.-H. Lin, Y.-J. Chen, C.-H. Wu, A. Y.-G. Fuh, J.-H. Liu, P.-C. Yang, Applied Physics Letters 2005, 86, 161120; c) H. Coles, S. Morris, Nature Photon 2010, 4, 676–685; d) S. Furumi, The Chemical Record 2010,

10, 394–408; e) J. Xiang, A. Varanytsia, F. Minkowski, D. A. Paterson, J. M. D. Storey, C. T. Imrie, O. D. Lavrentovich, P. Palffy-Muhoray, Proc. Natl. Acad. Sci. U.S.A. 2016, 113, 12925–12928; f) C. Yuan, W. Huang, Z. Zheng, B. Liu, H. K. Bisoyi, Y. Li, D. Shen, Y. Lu, Q. Li, Sci. Adv. 2019, 5, eaax9501; g) J. Mysliwiec, A. Szukalska, A. Szukalski, L. Sznitko, Nanophotonics 2021, 10, 2309–2346.

[7]     D.-H. Kim, A. D'Aléo, X.-K. Chen, A. D. S. Sandanayaka, D. Yao, L. Zhao, T. Komino, E. Zaborova, G. Canard, Y. Tsuchiya, E. Choi, J. W. Wu, F. Fages, J.-L. Brédas, J.-C. Ribierre, C. Adachi, Nature Photon 2018, 12, 98–104.

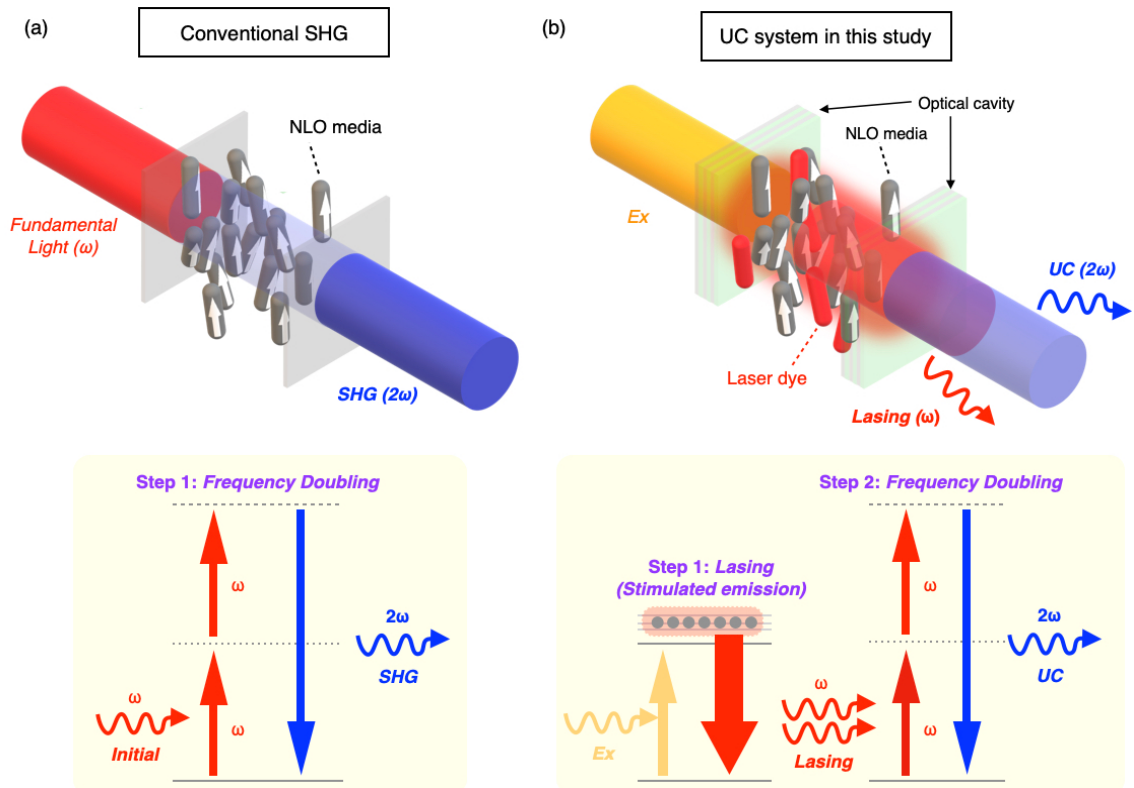

**Figure 1.** (a) The conventional up-conversion scheme of SHG. The propagating incident fundamental light photons are directly converted to the frequency-doubled SHG photons. (b) The up-conversion (UC) system demonstrated in the present paper, which is the simultaneous stimulated emission (lasing) and SHG processes. Here, the incident laser is used for excitation of fluorescent material inside cavity.

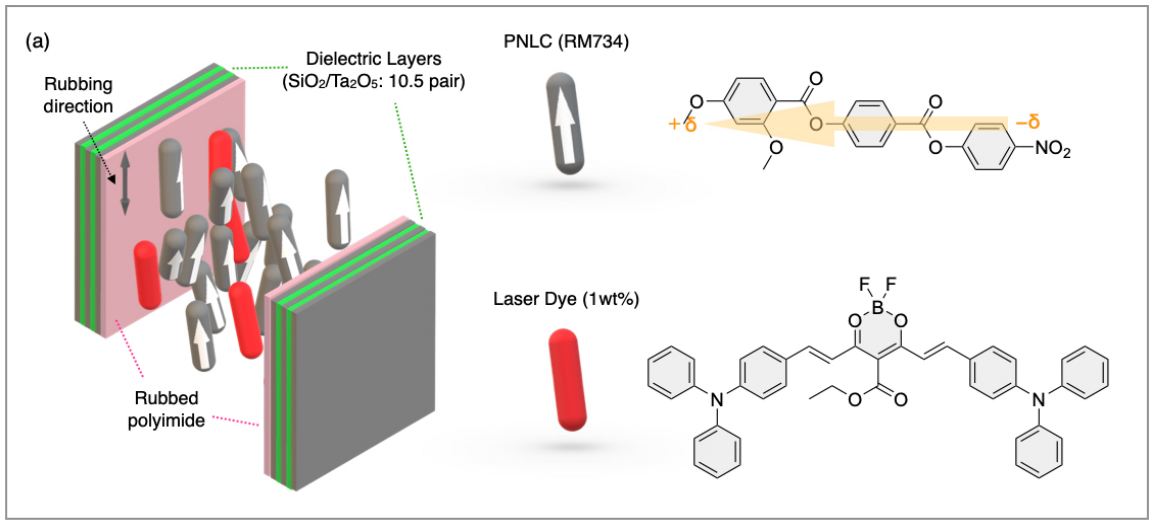

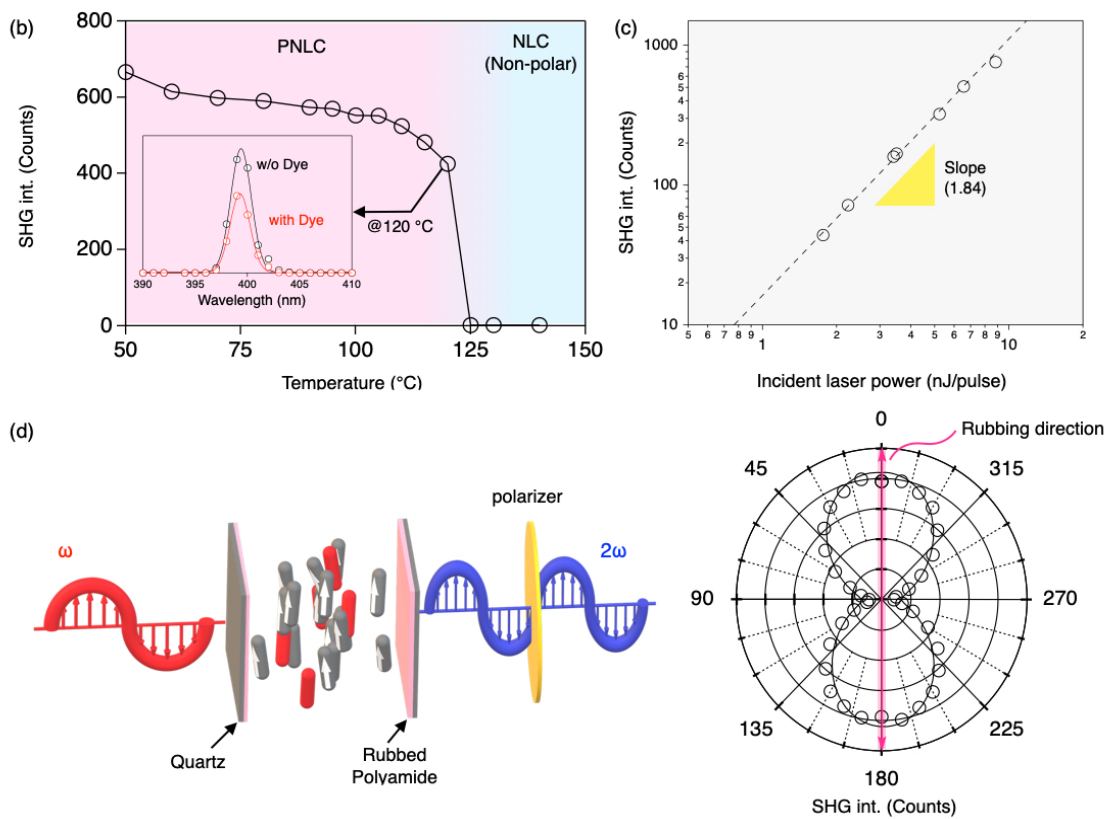

**Figure 2.** (a) The schematic representation of the optical cavity, and the molecular structures of the used PNLC (RM734) and fluorescent dye. (b) Temperature dependence of SHG intensity from PNLC. The SHG activity is confirmed below 120℃ at the phase transition temperature to the ferroelectric nematic state. Inset figure shows SHG spectra at 120℃ in doped (black) and undoped (red) PNLC with the fluorescent dye. (c) SHG intensity as a function of incident laser intensity. (d) Polar plot of the polarization angle dependence of the SHG signal.

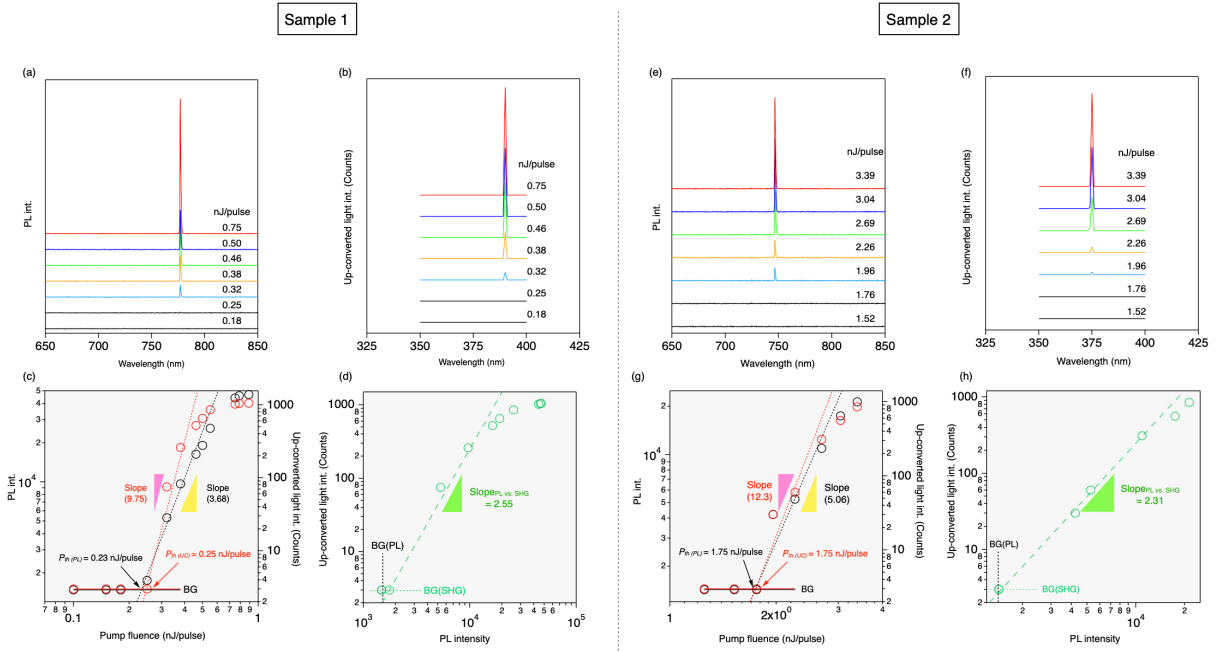

**Figure 3** (a, b, e, f) The PL and up-converted light spectrum at different excitation intensity in two different samples, in which lasing peaks appear at (a) 777 nm (Sample 1) and (e) 747 nm (Sample 2), respectively. (b, f) The up-converted light peaks appear at the half wavelengths of the lasing peaks. (c, g) The PL and up-converted light intensity plots as functions of the excitation (incident laser) intensities. The estimated thresholds ($P_{th}$ (PL), $P_{th}$ (UC)) are indicated together with their logarithmic slopes. (d, h) The up-converted light intensity potted as a function of the PL (lasing) intensity.

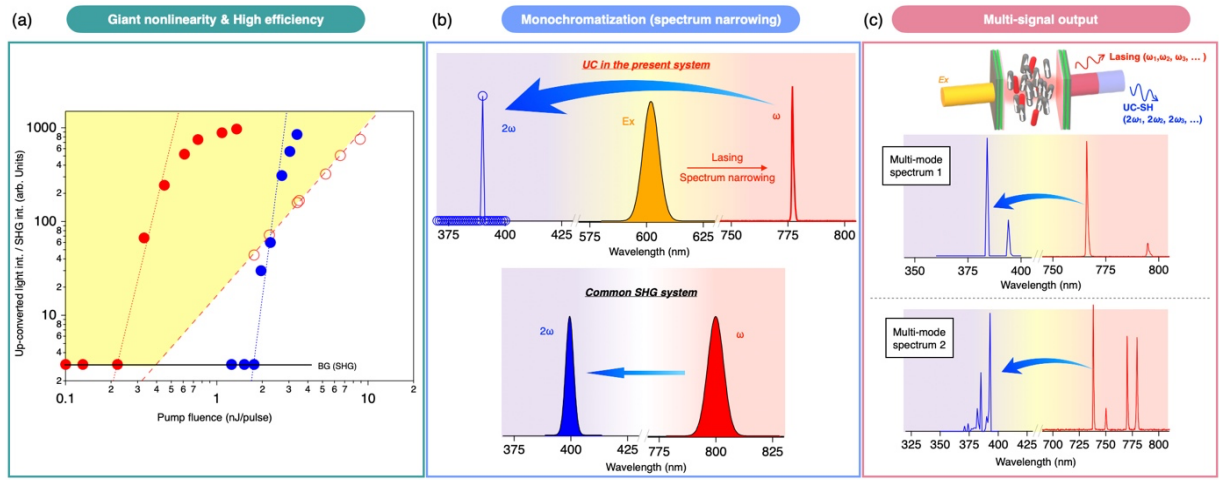

**Figure 4.** (a) Comparison of the intracavity up-conversion (Sample1, Sample2) and the conventional SHG (reprises of Figures. 3c, 3g, and 2d, respectively). The up-conversion process shows much more significant nonlinearity. The conversion efficiency is improved up to about two orders of magnitude higher than the conventional SHG process when the lasing threshold is lower (red filled circles in (a)). (b) Schematics of spectral narrowing in the present up-conversion process mediated by the stimulated emission. (c) Multi-mode coherent up-conversion.

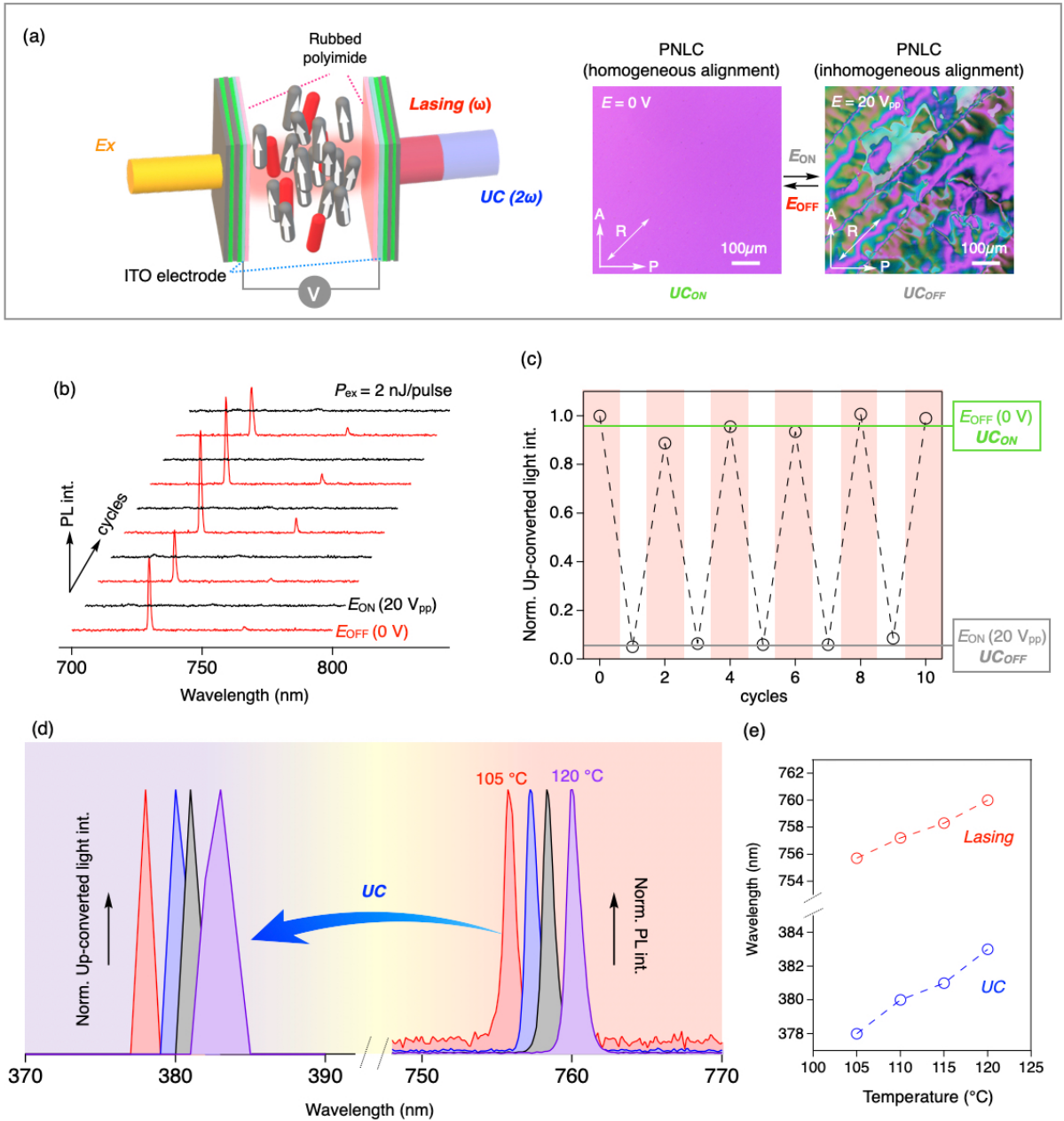

**Figure 5** (a) Schematic of the used F-P microcavity with transparent electrodes (ITO) and the polarized microscope images with and without electrical voltage. Here, the AC voltage (20Vpp, 200Hz) was applied to the device. (b, c) ON/OFF switching of (b) Lasing and (c) up-converted light under the electric-field operation. (d) Variation of lasing and up-conversion spectra upon cooling. (e) Their peak top wavelengths depending on the temperature.

# Supporting Information

Daichi Okada*, Hiroya Nishikawa, Fumito Araoka*

**Table of Contents**

# 1. Materials and Instruments

Unless otherwise noted, all reagents and solvents were used as received. The ferroelectric nematic liquid crystal (RM734) was purchased from Instec., Inc, USA. The used fluorescent dye was a commercially available boron difluoride curcuminoid derivative from Luminescence Technology Corp., Taiwan. The alignment material (AL1254) was a product of JSR Corp., Japan.

UV-VIS absorption and transmission spectra were recorded using a JASCO V-770 spectrophotometer. The dielectric mirror of alternating layers of 10.5 pairs of $SiO_2$ and $Ta_2O_5$ was fabricated using a magnetron sputtering system (SRV4320, Shinko-Seiki, Japan).

## 2. Linear optical response and sample preparation

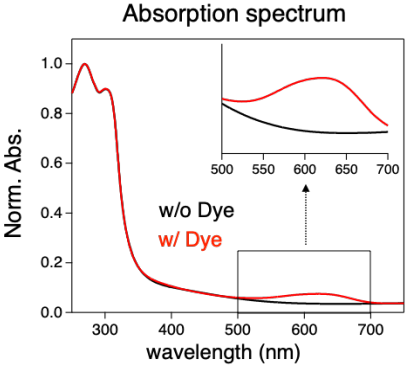

**Figure S1,** Absorption spectra of RM734 and a dye-doped RM734 thin films spin-coated on a quartz substrate.

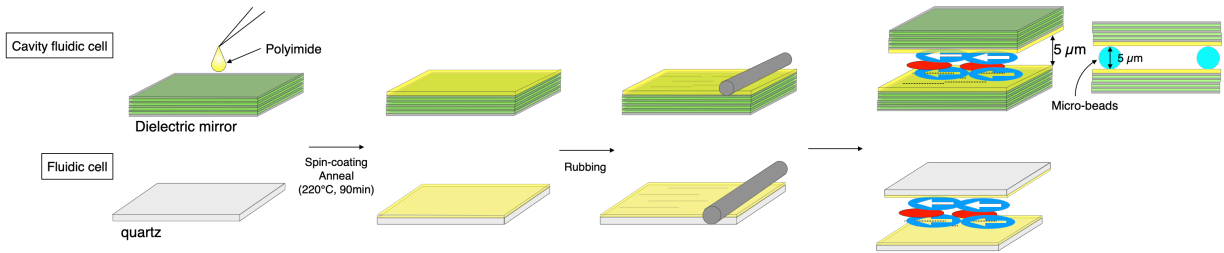

**Figure S2** Schematization of the fabrication process of the F-P microcavity device based on a LC cell.

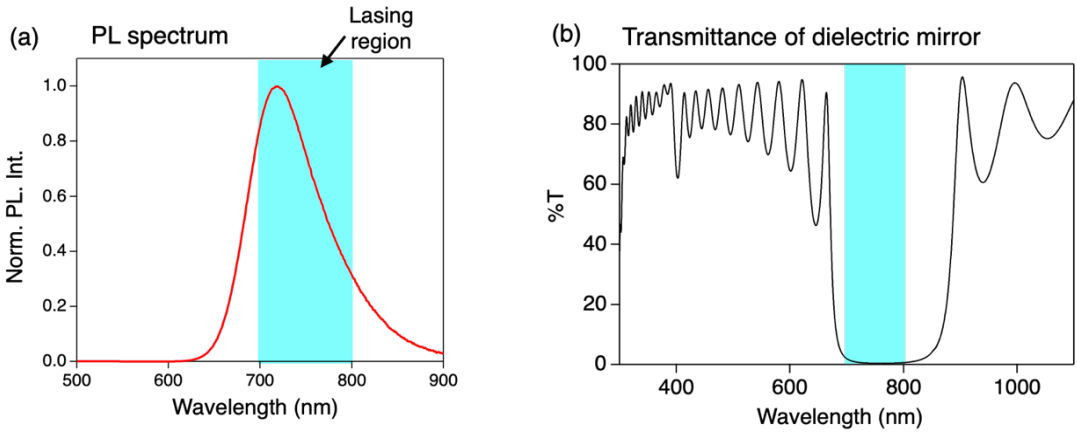

**Figure S3**, (a) The PL spectrum of the dye-doped PNLC; (b) Transmission of the dielectric mirror whose reflection band is adjusted to the PL peak wavelength (indicated by a light blue box), where the lasing action is highly probable.

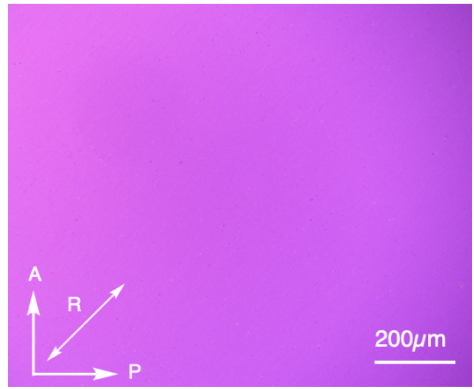

**Figure S4,** A polarized microscope image of the planarly-aligned dye-doped PNLC in a LC cell at 120°C.

## 3, Optical setup

Our optical setup is illustrated in Fig. S5. We mainly used two laser sources: one at 800 nm from regenerative amplifier and the other at 600nm from an optical parametric amplifier with a regenerative amplifier (OPA9400 with RegA, Coherent Inc., USA), both operating at a frequency of 10 kHz and with a pulse width of ~200 fs. The 600 nm pulse was used to excite the fluorescent dye, and the 800 nm is used for typical SHG investigation. The sample was heated in a hot stage with a temperature controller (HSC402 with mk2000, Instec Inc., USA). The obtained signal was collected by an objective lens (Mitsutoyo Corp., Japan) in the transmission direction. The PL spectra were recorded using an optical multichannel analyser (OMA) (USB4000, Ocean Optics Inc., USA), while SHG was detected with a photon-counting head (H7421, Hamamatsu Photonics K.K., Japan) on a monochromator. The SHG spectra were recorded in steps of 1 nm. Reference SHG spectra measured using these two spectroscopic setups were compared, and we confirmed that there is no need for calibration between them.

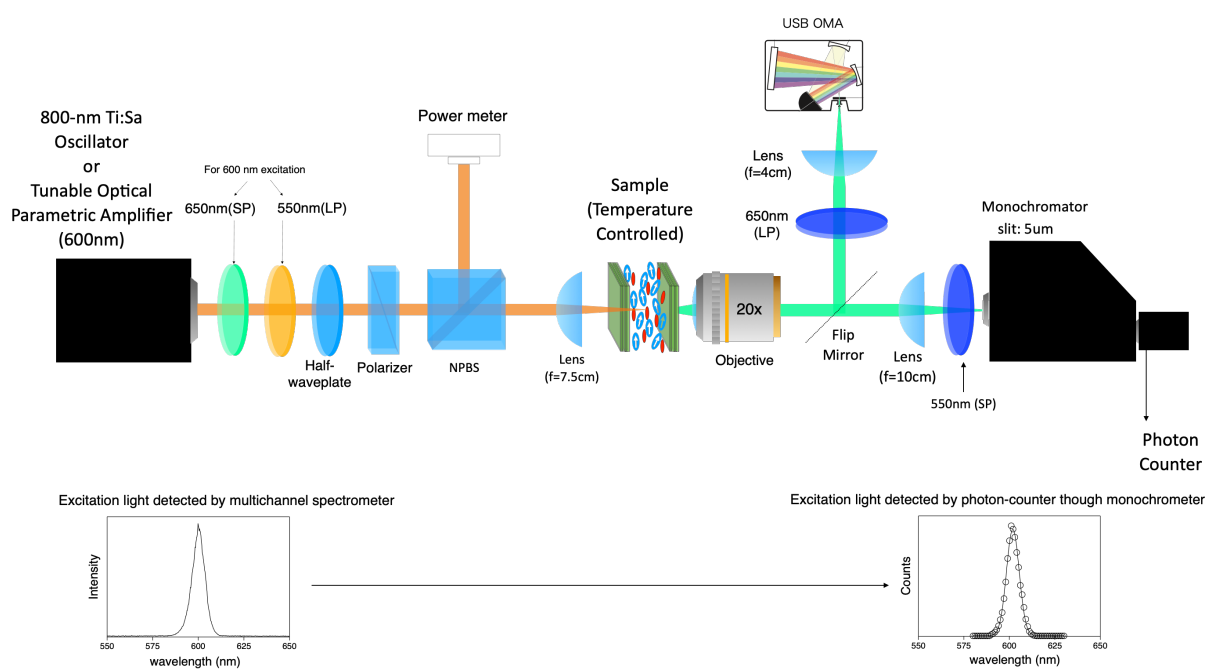

**Figure S5,** Our homemade optical setup for PL and SHG spectroscopy.

## 4, SHG and UC

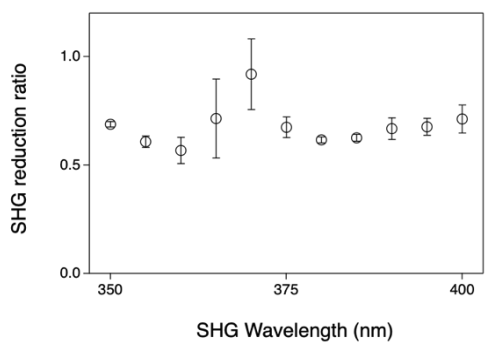

**Figure S6**, Influence of dye doping on SHG as a reduction ratio depending on the wavelength. The fundamental laser wavelength was scanned from 800 nm to 700 nm in steps of 10 nm, the reduction ratio was defined as a ratio of SHG intensities from the samples with and without dye-doping.

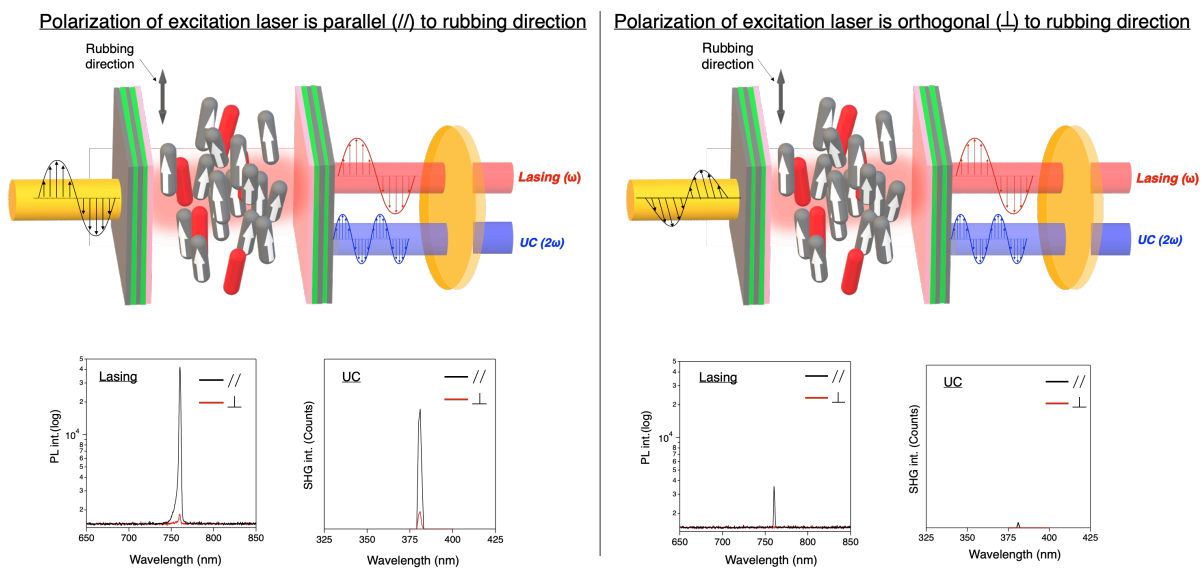

**Figure S7,** The polarization dependency of output optical signal under the different excitation polarization conditions.

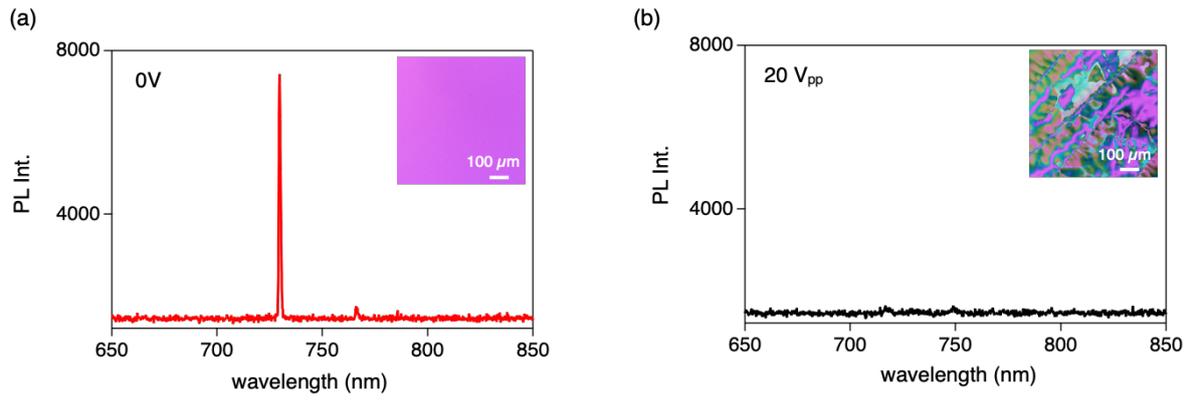

**Figure S8,** ON/OFF behavior of lasing (a) without and (b) with applying an electric field.